\documentclass{aa}
\usepackage{txfonts}

\usepackage{graphicx}
\usepackage{natbib}
\bibpunct{(}{)}{;}{a}{}{,}

\usepackage{multirow}
\begin{document}

\title{Methanol detection in \object{M\,82}}
\author{
Mart\'{\i}n, S. \inst{1} \and 
Mart\'{\i}n-Pintado, J. \inst{2} \and 
Mauersberger, R.\inst{1} 
}
\offprints{S.Mart\'{\i}n, \email{martin@iram.es}}
\institute{
Instituto de Radioastronom\'{\i}a Milim\'etrica, Avda. Divina Pastora 7, Local 20, 18012 Granada, Spain \and
Departamento de Astrof\'{\i}sica Molecular e Infrarroja, Instituto de Estructura de la Materia, CSIC, Serrano 121, 28006 Madrid, Spain
}
\date{Received <date> / Accepted <date>}

\abstract
{The nuclear starburst region in M\,82 shows systematical low abundances of some complex molecules when compared 
with other starburst galaxies.
This is likely related to a presumably photodissociation dominated environment.
In particular, methanol is known to show relatively low abundance because it is easily photodissociated.}
{We present a multilevel study of the emission of methanol, detected for the first time in this galaxy, and discuss the origin of its emission.}
{Observations of three transitions of CH$_3$OH towards the center and two positions around the nucleus of M\,82 are presented.
Two different components are found, one with  high excitation ($n(\rm H_2)\sim 10^6 cm^{-3}$, $T_{\rm rot}\sim 20$\,K) and a the
other with low excitation ($n(\rm H_2)\sim 10^4 cm^{-3}$, $T_{\rm rot}\sim 5$\,K).}
{The high observed methanol abundance of a few $10^{-9}$  can only be explained if injection of methanol from dust grains 
is taken into account. While the overall [CH$_3$OH]/[NH$_3$] ratio is much larger than observed towards other starbursts, the dense high 
excitation component shows a similar value to that found in NGC\,253 and Maffei\,2.
}
{Our observations suggest the molecular material in M\,82 to be formed by dense warm cores, shielded from the UV radiation and similar to the molecular
clouds in other starbursts, surrounded by a less dense photodissociated halo. The dense warm cores are likely the location of recent 
and future star formation within M\,82.}


\maketitle

\section{Introduction}
The nuclear environment of \object{M\,82} (\object{NGC\,3034}, \object{Arp\,337}, \object{3C231}) houses one of the brightest extragalactic IRAS sources 
\citep{Soifer89}
and represents an archetype of a nearby \citep[$D$=3.6\,Mpc,][]{Freedman94} nuclear starburst 
galaxy, where stars are forming at a rate of $\sim 9\,\rm M_\odot yr^{-1}$ \citep{Strickland04}.
The large supply of molecular gas within the inner few hundred parsecs has been extensively studied in CO and
other molecular tracers \citep[e.g.][]{Mao00}.
Abundances relative to H$_2$ of some complex molecules like SiO, CH$_3$OH, HNCO, CH$_3$CN and NH$_3$ are 
systematically low in comparison to similar starburst prototypes, e.g. \object{NGC\,253} \citep{Takano02,Martin06}.
These chemical differences have been interpreted as a consequence of 
nuclear starburst in \object{M\,82} being in a more evolved stage than that in \object{NGC\,253} \citep{Wang04}. 
Observations of HCO
\citep{Burillo02} suggest that photo-dissociation dominates the heating and the chemistry of the molecular material in the nuclear region of \object{M\,82}.

Methanol (CH$_3$OH) emission is observed in a wide range of physical conditions within the Galactic and 
extragalactic interstellar medium \citep{Menten88, Henkel87, Hutte97} and also in comets \citep{Bockelee91}.
The observed gas-phase abundances range from a few $\sim10^{-9}$ in dark clouds \citep{Ohishi} and photo-dissociation 
regions (PDRs) \citep{Jansen95} 
reaching up to $\sim10^{-6}$ in hot molecular cores \citep{Sutton95,Nummelin} and Galactic Center molecular clouds 
(Requena-Torres et al. 2006, in prep).
CH$_3$OH is, after H$_2$O, the most abundant known constituent of interstellar ices \citep{Allam92}.
Observed solid-state CH$_3$OH abundances are 1 to 4 order of magnitude larger than those in gas-phase \citep{Ehren00, Schoier02}.
This fact, as well as the difficulties of gas-phase chemical models to produce fractional
abundances larger than a few $10^{-9}$ \citep{Lee96}, supports the assumption that large abundances of methanol are due to its 
injection into the gas-phase via evaporation and/or disruption of ice mantles \citep{Millar91,Charnley95}.

Since methanol is easily photodissociated by UV radiation as observed in galactic PDRs \citep{Hartquist95,Leteuff00}, 
one would expect that \object{M\,82} will show very low abundances.
In this letter we present first detections of CH$_3$OH towards the nuclear region of \object{M\,82}.
The fractional abundance of CH$_3$OH, detected at three positions, is larger than expected if the molecular composition in the nucleus of 
\object{M\,82} is exclusively dominated by PDRs. 
This fact indicates the presence of a significant fraction of dense gas well shielded from the UV radiation. 

\section{Observations and results}
The observations were carried out in July 
and December 
2004 with the IRAM 30\,m telescope on Pico Veleta, Spain.
Three methanol lines at 96.7, 157.2 and 241.7\,GHz were observed. 
The beam sizes of the telescope at these frequencies are $25'', 16''$ and $10''$, respectively.
Observations were made in symmetrical wobbler switching mode with a throw of $4'$ in azimuth and a frequency of 0.5\,Hz. 
We used the $512\times1$MHz filter bank for the 3\,mm and two $256\times4$\,MHz filter banks for the 2 and 1\,mm lines. 
The spectra were calibrated using a dual load system.
System temperatures in $T_{\rm MB}$ were $\sim 200$\,K (3\,mm), 370\,K (2\,mm) and 630\,K\,(1\,mm).
The nominal pointing position was $\alpha_{J2000}=09^{\rm h}55^{\rm m}51\fs9,\delta_{J2000}=69^{\rm o}40'47\farcs1$, 
corresponding approximately to the 2.2$\mu$m emission peak \citep{Dietz86, Joy87}.
Continuum cross-scans of point sources made every 2 hours led to a pointing accuracy of $\sim 3''$.
The central and two offset positions were observed.
The North-East (NE at $+13'',+7.5''$) and South-West (SW at $-13'',-7.5''$) positions correspond approximately to the HCO 
emission peaks seen in the interferometric maps by \citet{Burillo02}. 

\begin{figure}
\begin{center}
	\includegraphics[width=0.56\linewidth, angle=-90]{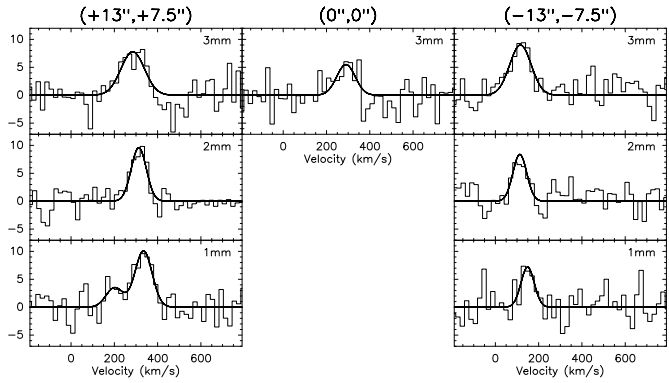}
\caption {Observed methanol lines towards three positions in the nucleus of \object{M\,82}. The spectra, in $T_{\rm MB}$\,(mK) scale,
have been smoothed to a 20\,km\,s$^{-1}$ velocity resolution.}
\label{fig:CH3OHlines}
\end{center}
\end{figure}

\begin{table}
\caption{Parameters derived from Gaussian fits to the observed lines.}
\label{tab:CH3OHfits}
\centering
\scriptsize
\begin{tabular}{c c r@{$\pm$}l r@{$\pm$}l r@{$\pm$}l r r }
\hline\hline
CH$_3$OH              &  Frequency\,$^a$ & \multicolumn{2}{c}{$\int{T_{\rm MB}{\rm d}v}$} & \multicolumn{2}{c}{$V_{\rm LSR}$} &  \multicolumn{2}{c}{$\Delta v_{1/2}$} & $T_{\rm MB}$  & $rms\,^b$ \\
                      &  (GHz)           & \multicolumn{2}{c}{mK\,km\,s$^{-1}$}           & \multicolumn{2}{c}{km\,s$^{-1}$}  &  \multicolumn{2}{c}{km\,s$^{-1}$}     & mK            & mK           \\
\hline
\multicolumn{4}{r}{\bf NE Lobe} & \multicolumn{6}{l}{$\bf (+13'',+7.5'')$}      \\               
$2_k-1_k$             &  96.7            & 1070 & 170                                     & 285 & 11                          &  128 & 19                             &  7.8          & 2.5 \\
$J_0-J_{-1}$          & 157.3            &  850 &  70                                     & 313 &  4                          &   82 &  8                             &  9.6          & 1.2 \\
$5_k-4_k$             & 241.8            & 1010 & 120                                     & 335 &  6                          &   93 &  6                             & 10.1          & 2.2 \\
                      & 241.9            &  350 & 100                                     & 310 & 10                          &   93 &  0                             &  3.5          & 2.2 \\
\multicolumn{4}{r}{\bf Center}  & \multicolumn{6}{l}{$\bf (0'',0'')$}                                \\    
$2_k-1_k$             &  96.7            &  580 & 190                                     & 290 & 20                          &  100 & 44                             &  5.5          & 2.6 \\
\multicolumn{4}{r}{\bf SW Lobe} & \multicolumn{6}{l}{$\bf (-13'',-7.5'')$}                           \\  
$2_k-1_k$             &  96.7            & 1120 & 180                                     & 116 &  8                          &  117 & 24                             &  9.0          & 2.5 \\
$J_0-J_{-1}$          & 157.3            &  670 & 110                                     & 113 &  6                          &   75 & 13                             &  8.4          & 2.0 \\
$5_k-4_k$             & 241.8            &  550 & 120                                     & 149 &  8                          &   72 & 19                             &  7.2          & 2.6 \\
\hline
\multicolumn{10}{l}{$^a$ approximate frequency of each group of transitions, used for velocity scale in Fig.~\ref{fig:CH3OHlines}.}\\
\multicolumn{10}{l}{$^b$ rms of the data at a 20\,km\,s$^{-1}$ velocity resolution.}\\
\end{tabular}
\end{table}

Fig.~\ref{fig:CH3OHlines} shows the observed CH$_3$OH line profiles, and Table~\ref{tab:CH3OHfits}
summarizes the parameters derived from the Gaussian fits to the observed lines in each position.
Towards the central position, the emission peaks at 290\,km\,s$^{-1}$ suggesting that most of the emission in the 
$25''$ beam is dominated by the NE clump.
Towards the other observed positions the linewidth of the 3\,mm transitions are broader than those of the 2 and 1\,mm lines.
This difference is likely due to the larger beam size at 3\,mm detecting a more extended component.

The seemingly low radial velocity of the 3\,mm transition towards the NE position is 
due to a Gaussian fit to a 
non-Gaussian profile. 
In fact, the emission peaks at higher velocities than that of the fit.
On the other hand, the systematically higher velocities shown by the 1\,mm transitions at 241.7\,GHz are due to the 
overlap of the different methanol transitions.

To estimate the size of the emitting region, we have followed the procedure used by \citet{Mauers03}.
From 
the CS $J=3-2$, C$^{34}$S $J=2-1$ and
$J=3-2$ lines we estimate an equivalent source size for the SW lobe of $\theta_s\sim12''$. 
A mean source size of $12''$ will be used for all three positions to convert the methanol main beam brightness temperatures
in Table~\ref{tab:CH3OHfits} into source averaged methanol column densities.
This size agrees with the extent of the NE and SW lobes seen in the high resolution HCO maps \citep{Burillo02}.

\begin{figure}
	\includegraphics[width=0.38\linewidth, angle=-90]{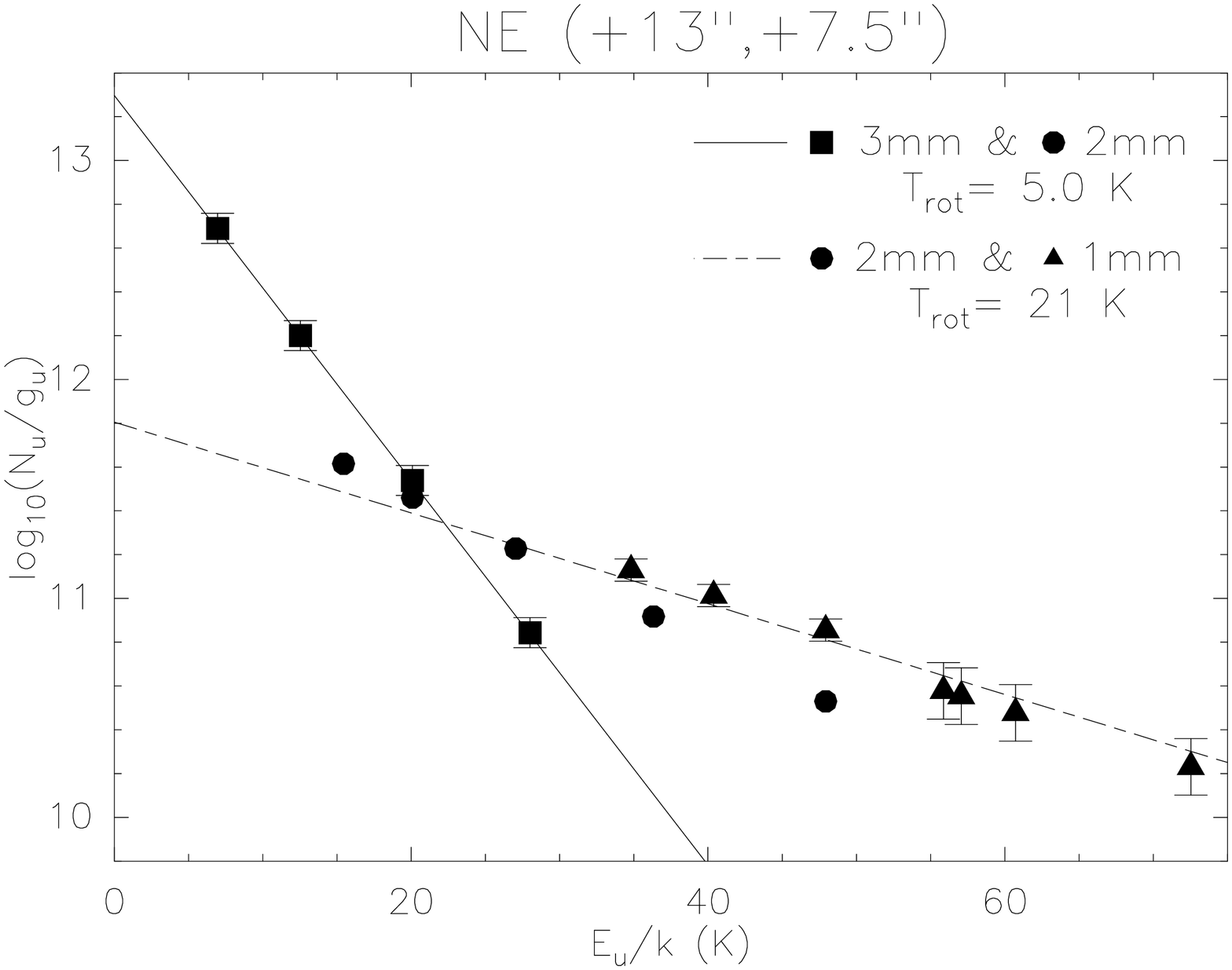}
	\hspace{-5.4 pt}
	\includegraphics[width=0.38\linewidth, angle=-90]{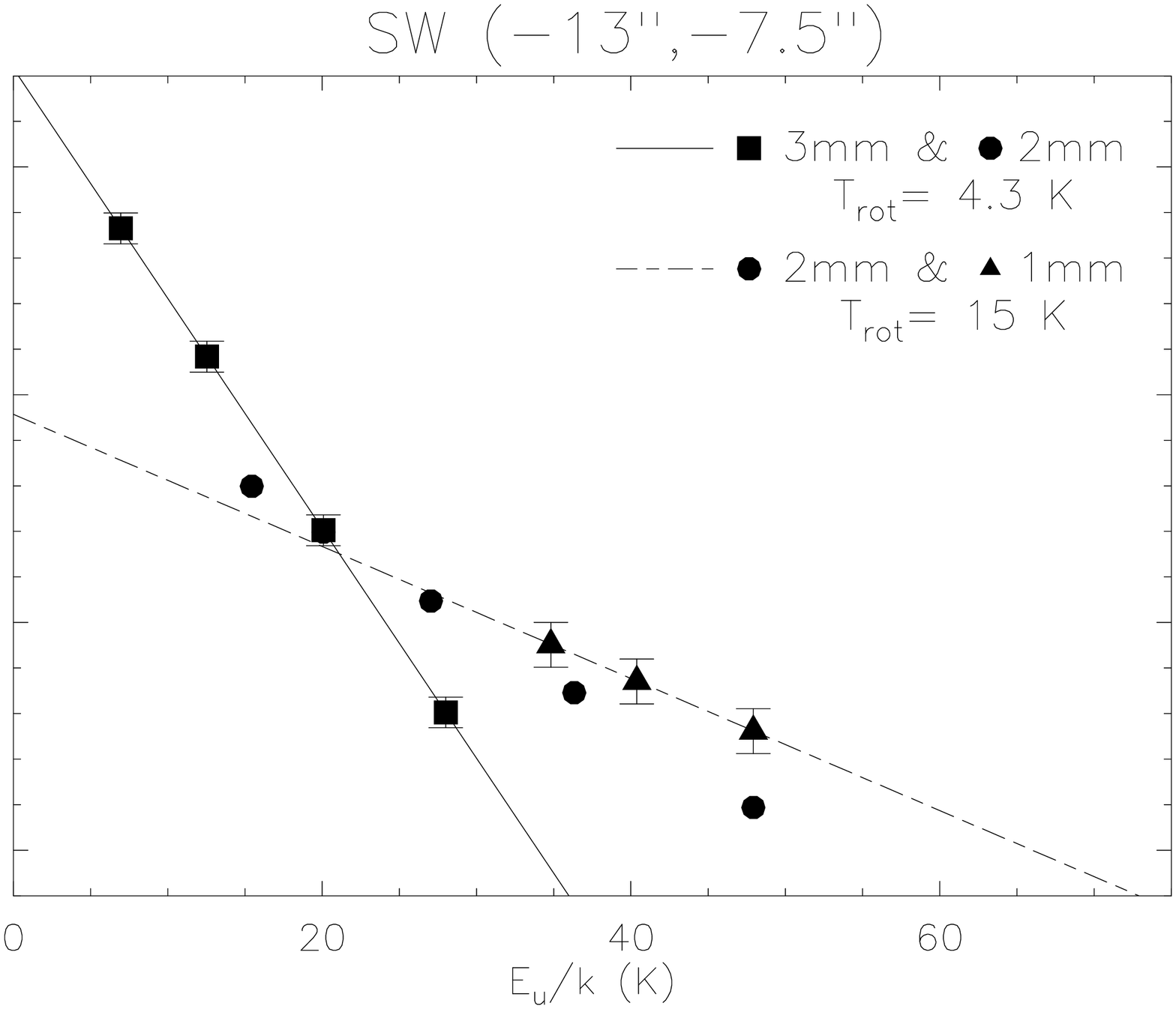}
\caption {Rotational diagrams for the the NE and SW positions.}
\label{fig:diagrot}
\end{figure}

Fig.~\ref{fig:diagrot} shows population diagrams for the NE and SW lobes corrected for beam dilution. 
Diagrams were plotted by estimating the contribution of each of the transitions to the whole
observed profiles \citep[see][]{Martin06}.
Methanol emission has been fitted with a two different temperature components.
The fit to the 3 and 2\,mm transitions appear to trace a 
low excitation gas with a rotational temperature ($T_{\rm rot}$) of $4-5$\,K and column densities of 
$N[\rm CH_3OH]=3\times10^{14} cm^{-2}$, similar towards both lobes. 
Towards the central position, where a $T_{\rm rot}\sim 4.5$\,K has been assumed, we derive a CH$_3$OH column 
density a factor of 2 smaller than towards the other positions.
In addition to the low excitation temperature component, 
the 2 and 1\,mm transitions trace
a much higher temperature component with $T_{\rm rot}=15$ and 21\,K, derived for the SW and NE lobes respectively.
The NE lobe shows a significantly higher rotational temperature than the SW lobe. Our data clearly show that the methanol emission in M82 samples two 
components with very different (low and high) excitation conditions. This indicates the presence of large gradients in the excitation conditions in the 
molecular gas where methanol emission arises. Assuming a kinetic temperature, $T_{\rm kin}\sim 30$\,K similar to that derived from NH$_3$ observations 
\citep{Weiss01}, we find that a large fraction of methanol arises from a relatively low density gas with $\rm n_{H_2}\sim10^{4}cm^{-3}$ and a smaller 
fraction from denser gas with $\rm n_{H_2}>10^{6}cm^{-3}$.  
The derived H$_2$ densities do not vary significantly if we assume a $T_{\rm kin}\sim 100$\,K as derived for the warmer component observed in CO \citep{Mao00}. 

To estimate the CH$_3$OH fractional abundance, we have calculated the column density of molecular hydrogen, $N({\rm H_2})$, from the $^{13}$CO\,$J=2-1$ 
observations by \citet{Mao00} 
using the conversion $N({\rm H_2})=3.3\times10^{20}{\rm cm}^{-2}I({\rm ^{13}CO_{2-1}})$ from \citet{Mauers03}.
Methanol fractional abundances for the low and high excitation components in Table~\ref{tab:CH3OHdensities}
are one order of magnitude higher than the upper abundance limit by \citet{Hutte97}.
This disagreement is due to the different H$_2$ column density and the beam filling factor correction
applied in this work.
The methanol abundance estimate of a few $10^{-9}$ in M\,82 is still significantly lower than that of $10^{-7.3}-10^{-7.9}$ 
in other starburst galaxies \citep{Martin06}.

\begin{table}
\caption{Physical parameters derived from observations.}
\label{tab:CH3OHdensities}
\centering
\scriptsize
\begin{tabular}{c @{~} c @{~} c @{~} c @{~} c @{~} c @{~} c }
\hline\hline
Position                       &     ${\rm ^{13}CO}_{J=2-1}\,^{a}$   &   $N(\rm H_2) ^{b}$     &  $N[\rm CH_3OH] ^{b}$  &   $T_{\rm rot}$  &     $X$[CH$_3$OH]      &   n(H$_2$)$^{c}$           \\    
($''$,$''$)                    &       (K\,km\,s$^{-1}$)           & $10^{22}{\rm cm}^{-2}$  &  $10^{13}{\rm cm}^{-2}$  &      (K)         &       (10$^{-9}$)      &     (cm$^{-3}$)            \\     
\hline
\multirow{2}{*}{(+13,+7.5)}    & \multirow{2}{*}{58.9}             &  \multirow{2}{*}{8.5}   &  27                     &  $5.0\pm0.1$       &         3.2           &     $5.3-5.4\times10^4$   \\     
                               &                                   &                         &  7.5                    &  $21\pm 1$         &         0.9           &     $5.7-1.6\times10^6$   \\     
(0,0)                          &         59.8                      &   8.6                   &  15                     &  $4.5      $       &         1.7           &                           \\     
\multirow{2}{*}{(-13,-7.5)}    & \multirow{2}{*}{69.0}             &  \multirow{2}{*}{9.9}   &   29                    &  $4.3\pm0.1$       &         2.9           &     $4.1-4.2\times10^4$   \\     
                               &                                   &                         &   5.9                   &  $15\pm 1$         &         0.6           &     $2.2-1.1\times10^6$   \\     
\hline
\multicolumn{7}{l}{$^{a}$ Integrated intensity from \citet{Mao00} for a beamwidth of $22''$.} \\
\multicolumn{7}{l}{$^{b}$ Source averaged column density assuming an emission extent of $12''$.} \\
\multicolumn{7}{l}{$^{c}$ Assuming a $T_{\rm kin}=30-100$\,K, respectively.} \\
\multicolumn{7}{l}{~~Collisional coefficients from \citep{Pottage04}.}  \\
\end{tabular}
\end{table}

\section{Discussion}

\subsection{The origin of CH$_3$OH in \object{M\,82}: Photodissociation of core-halo molecular clouds}

The chemical composition of \object{M\,82} shows a clear resemblance with that observed in the \object{Orion Bar}, 
prototype of Galactic PDR \citep{Martin06},
supporting the idea that the chemistry of the nuclear molecular environment of \object{M\,82} 
could be dominated by photo-dissociating radiation \citep{Burillo02, Mauers03}. 
This is considered to be the signature of a more evolved stage of its nuclear starburst than in other nearby 
starburst galaxies such as \object{NGC\,253} and \object{NGC\,4945} \citep{Wang04}. 

The derived overall fractional abundance of methanol, taking into account the two components (low and high excitation),
of  $\sim 4\times10^{-9}$, similar towards both positions (see Table~\ref{tab:CH3OHdensities}),
is about a factor $5-10$ larger than the values of $0.5-1\times10^{-9}$ found in Galactic PDRs like the \object{Orion Bar}
\citep{Jansen95,Johnstone03}.
This suggests that, in addition to photo-dissociation, another type of chemistry is required to explain the 
high methanol abundances measured in \object{M\,82}.

Even the most favorable predictions of gas-phase chemistry models cannot reproduce
methanol abundances larger than a few $10^{-9}$ \citep{Lee96}.
Only models which take into account methanol injection from icy mantles into gas-phase are able to 
produce large scale high abundances \citep{Millar91,Charnley95}. 
Thermal evaporation of ices in the nuclear environment of \object{M\,82} is unlikely given the low global dust temperature 
\citep[$T_{\rm d}\sim 50$\,K,][]{Negishi01}, which is below the typical critical temperature above which icy mantles 
evaporation becomes important \citep[$T_{\rm crit}>100$\,K,][]{Isobe70,Collings04}
Like in the nuclei of  \object{NGC\,253} and the \object{Milky Way},
low velocity shocks within molecular cloud complexes seem to be the most likely way of injecting alcohols into the gas-phase \citep{Pintado01,Requena06}.

However, the abundance of ammonia, also ejected from icy grain mantles \citep{Flower95} 
and easily photodissociated like methanol \citep{Fuente90},
is much lower in \object{M\,82} than in \object{NGC\,253} and \object{Maffei\,2}. 
Table~\ref{tab:CH3OHtoNH3} shows the comparison of the CH$_3$OH and NH$_3$ column densities in \object{M\,82} SW lobe with those observed in the starburst galaxies 
\object{NGC\,253} and \object{Maffei\,2}. 
While the total ratio [CH$_3$OH]/[NH$_3$]  is similar in \object{NGC\,253} and \object{Maffei\,2}, the overall abundances ratio in \object{M\,82}  is a factor of 5 larger than 
in the other galaxies. 
It is interesting to note that if one considers that all ammonia in \object{M\,82} arises from the high density component the [CH$_3$OH]/[NH$_3$] abundance ratio is similar 
to that measured in \object{NGC\,253} and \object{Maffei\,2}. 

The emerging scenario from our methanol observations is that the molecular clouds in \object{M\,82} have a core-halo structure with a large
density contrast.
The UV radiation penetrates into the halos, photodissotiating both NH$_3$ and CH$_3$OH.
We find a large methanol to ammonia abundance ratio in the halo as a result of a high temperature chemistry and the slightly different photodissociation 
cross-sections between both molecules \citep{Leteuff00}. 
On the other hand, in the high density cores, well shielded from the UV radiation,
NH$_3$ and CH$_3$OH abundances 
reflect the typical abundance ratio found in other starburst galaxies.

\citet{Viti02} have followed the complete chemical evolution of a collapsing clump affected by slow shocks and photodissociating UV radiation.
In the collapsing phase, molecules which are formed in gas phase freeze out onto the surface of dust grains. 
Afterwards, slow shocks eject 
molecules from the grains which are also photodissociated by the UV radiation. 
Although the conditions in the collapsing core are different than in the molecular clouds in the  nucleus of \object{M\,82}, the model gives insights on the 
evolution of methanol and ammonia abundances from the shielded cores dominated by the ejection from grains to the low density halo dominated by 
photochemistry.
In their high temperature model \citep[see Fig.4 from][]{Viti02}, after the ejection of molecules from grains by shocks, the cores 
show ammonia abundances larger than those of methanol by more than one order the magnitude, similar to what is observed in NGC\,253 
and the dense warm cores of \object{M\,82}. 
As the UV radiation impinges the gas, the [CH$_3$OH]/[NH$_3$] ratio increases and after about 100 years the methanol abundance 
is an order of magnitude larger than that of ammonia. 
The trend predicted by the 
model in the abundance ratios is consistent with the idea that the molecular clouds in \object{M\,82} 
are composed of high density cores surrounded by
relatively large low density envelopes whose chemistry is dominated by photochemistry. 
In fact, \citet{Mao00} also suggested that the bulk of the CO emission from the nuclear region in \object{M\,82} stems from a
UV-illuminated low density fragmented interclump medium likely associated with the low excitation component observed in methanol.

\subsection{The large scale distribution of methanol in \object{M\,82}}
Recent multiline high-resolution maps towards \object{IC\,342} \citep{Meier05} show a clear difference in the location between the PDRs 
towards the inner nuclear region and large-scale shocks observed along the nuclear arms.
The picture of the central parsec of \object{M\,82} appears to be fairly different. 
The emission of methanol seems to follow the double-lobed structure observed in all molecules, 
which has been interpreted as a molecular ring \citep{Weli84,Nakai87} with a gas depletion in the central
region.
Observation of molecules claimed to trace PDR show an appreciable emission toward the central region besides the double-peak structure 
\citep{Burillo02}.
Methanol emission in \object{M\,82}, on the other hand, appears almost fully concentrated toward the lobes at both sides of the nucleus as
the profile observed in the central position can be explained from the emission arising mainly from the NE concentration.


Finally, if the methanol emission arising from the densest regions in M\,82, shielded from the UV radiation, represents the location of recent 
and/or ongoing star formation, the low column densities measured in this component suggest that \object{M\,82} will have less high density gas available 
for future star formation. This is in contrast with other starburst galaxies like \object{NGC\,253} and \object{Maffei\,2} and in agreement with 
\object{M\,82} being in a more evolved stage.

\begin{table}
\caption{Methanol to Amonia ratio in starbursts.}
\centering
\scriptsize 
\begin{tabular}{c c c c c c}
\hline\hline
                           &  \multicolumn{3}{c}{\object{M\,82} $^a$}  &  \object{NGC\,253} $^b$ &   \object{Maffei\,2}  \\
                           & Total     &  Halo $^c$     &  Core $^c$   &               &              \\
\hline  
  N[CH$_3$OH]              &  34.9     &   29          &   5.9        &   83          &      33      \\
  N[NH$_3$] $^d$           &  36       & $\ll36$       & $\la  36$    &  440          &     140      \\
  $\rm [CH_3OH]/[NH_3]$    &  0.97     & $\gg0.81$     & $\ga  0.16$  &  0.19         &    0.24      \\
\hline
\multicolumn{6}{l}{Column densities in units of $10^{13}{\rm cm}^{-2}$.} \\
\multicolumn{6}{l}{$^a$ Data for the SW lobe.}\\
\multicolumn{6}{l}{$^b$ Both lobes included.}\\
\multicolumn{6}{l}{$^c$ Assuming most of the NH$_3$ arises from the high density component.}\\
\multicolumn{6}{l}{$^d$ Data from \citet{Weiss01,Mauers03}.}\\
\end{tabular}
\label{tab:CH3OHtoNH3}
\end{table}

\begin{acknowledgements}
JMP has been partially supported by the Spanish Ministerio de Educaci\'on y Ciencia under projects ESP~2004-00665, AYA 2002-10113-E and AYA~2003-02785-E.
\end{acknowledgements}


\bibliographystyle{aa}

\begin{thebibliography}{}
\bibitem[Allamandola et al.(1992)]{Allam92} Allamandola, L.J., Sandford, S.A., Tielens, A.G.G.M., \& Herbst, T.M. 1992, ApJ, 399, 134
\bibitem[Bockel\'ee-Morvan et al. (1991)]{Bockelee91} Bockel\'ee-Morvan, D., Colom, P., Crovisier, J., Despois, D., \& Paubert, G. 1991, Nature, 350, 318
\bibitem[Charnley et al.(1995)]{Charnley95} Charnley, S.B., Kress, M.E., Tielens, A.G.G.M., \& Millar, T.J. 1995, ApJ, 448, 232
\bibitem[Collings et al.(2004)]{Collings04} Collings, M.P., Anderson, M.A., Chen, R. et al. 2004, MNRAS, 354, 1133
\bibitem[Dietz et al.(1986)]{Dietz86} Dietz, R.D., Smith, J., Hackwell, J.A., Gehrz, R.D., \& Grasdalen, G.L. 1986, AJ, 91, 758
\bibitem[Ehrenfreund \& Charnley(2000)]{Ehren00} Ehrenfreund, P., \& Charnley, S.B. 2000, ARA\&A, 38, 427
\bibitem[Flower et al.(1995)]{Flower95} Flower, D.R., Pineau des For\^ets, G., \& Walmsley, C.M. 1995, A\&A, 294, 815
\bibitem[Freedman et al.(1994)]{Freedman94} Freedman, W.L., Hughes, S.M., Madore, B.F., et al. 1994, ApJ, 427, 628
\bibitem[Fuente et al.(1990)]{Fuente90} Fuente, A., Mart\'{\i}n-Pintado, J., Bachiller, R., \& Cernicharo, J. 1990, A\&A, 237, 471
\bibitem[Fuente et al.(2005)]{Fuente05} Fuente, A., Garc\'{\i}a-Burillo, S., Gerin, M., et al. 2005, ApJ, 619, L158
\bibitem[Garc\'{\i}a-Burillo et al.(2002)]{Burillo02} Garc\'{\i}a-Burillo, S., Mart\'{\i}n-Pintado, J., Fuente, A., \& Usero, A. 2002, ApJ, 575, L55
\bibitem[Hartquist et al.(1995)]{Hartquist95} Hartquist, T.W., Menten, K.M., Lepp, S., \& Dalgarno, A. 1995, MNRAS, 272, 184
\bibitem[Henkel et al.(1987)]{Henkel87} Henkel, C., Jacq, T., Mauersberger, R., Menten, K.M., \& Steppe, H. 1987, A\&A, 188, L1-L4
\bibitem[H\"uttemeister et al.(1997)]{Hutte97} H\"uttemeister, S., Mauersberger, R., \& Henkel, C. 1997, A\&A, 326, 59
\bibitem[Isobe(1970)]{Isobe70} Isobe, S. 1970, PASJ, 22, 429
\bibitem[Jansen et al.(1995)]{Jansen95} Jansen, D.J., Spaans, M., Hogerheijde, M.R., \& van Dishoeck, E.F. 1995, A\&A, 303, 541
\bibitem[Johnstone et.al(2003)]{Johnstone03} Johnstone, D., Boonman, A.M.S., \& van Dishoeck, E.F. 2003, A\&A, 412, 157
\bibitem[Joy et al.(1987)]{Joy87} Joy, M., Lester, D.F., \& Harvey, P.M. 1987, ApJ, 319, 314
\bibitem[Lee et al.(1996)]{Lee96} Lee, H.-H., Bettens, R.P.A., \& Herbst E. 1996, A\&AS, 119, 111
\bibitem[Le Teuff et al.(2000)]{Leteuff00} Le Teuff, Y.H., Millar, T.J., \& Markwick, A.J. 2000, A\&AS, 146, 157
\bibitem[Mao et al.(2000)]{Mao00} Mao, R.Q., Henkel, C., Schulz, A., et al. 2000, A\&A, 358, 433 
\bibitem[Mart\'{\i}n et al.(2006)]{Martin06} Mart\'{\i}n, S., Mauersberger, R., Mart\'{\i}n-Pintado, J., Henkel, C., \& Garc\'{\i}a-Burillo, S. 2006, ApJS, In Press
\bibitem[Mart\'{\i}n-Pintado et al.(2001)]{Pintado01} Mart\'{\i}n-Pintado, J., Rizzo, J.R., de Vicente, P., Rodr\'{\i}guez-Fern\'andez, N.J., \& Fuente, A. 2001, ApJ, 548, L65
\bibitem[Mauersberger et al.(2003)]{Mauers03} Mauersberger, R., Henkel, C., Wei\ss, A., Peck, A.B., \& Hagiwara, Y. 2003, A\&A, 403, 561
\bibitem[Menten et al.(1988)]{Menten88} Menten, K.M., Walmsley, C.M., Henkel, C., \& Wilson, T.L. 1988, A\&A, 198, 253
\bibitem[Meier \& Turner (2005)]{Meier05} Meier, D.S., \& Turner, J.L. 2005, ApJ, 618, 259
\bibitem[Millar et al.(1991)]{Millar91} Millar, T.J., Herbst, E., \& Charnley, S.B. 1991, ApJ, 369, 147
\bibitem[Nakai et al.(1987)]{Nakai87} Nakai, N., Hayashi, M., Handa, T., Sofue, Y., \& Hasegawa, T. 1987, PASJ, 39, 685
\bibitem[Negishi et al.(2001)]{Negishi01} Negishi, T., Onaka, T., Chan, K.-W. \& Roellig, T. L. 2001, A\&A, 375, 566
\bibitem[Nummelin et al.(2000)]{Nummelin} Nummelin A.,  Bergman P., Hjalmarson \AA., et al. 2000, ApJS, 128, 213
\bibitem[Ohishi et al.(1992)]{Ohishi} Ohishi, M., Irvine, W. M., \& Kaifu, N. 1992, in IAU Symp. 150, Astrochemistry of Cosmic Phenomena, ed. P.D. Singh (Dordrecht: Kluwer), 171
\bibitem[Pottage et al.(2004)]{Pottage04} Pottage, J.T., Flower, D.R., \& Davis, S.L. 2004, MNRAS, 352, 39
\bibitem[Requena-Torres et al.(2006)]{Requena06} Requena-Torres, M.A., Mart\'{\i}n-Pintado, J., Rodr\'{\i}guez-Franco, A, et al. 2006, Submitted
\bibitem[Sandford \& Allamandola(1993)]{Sandford93} Sandford, S.A. \& Allamandola, L.J. 1993, ApJ, 417, 815
\bibitem[Sch\"oier et al.(2002)]{Schoier02} Sch\"oier, F.L., J\o rgensen, J.K., van Dishoeck, E.F., \& Blake, G.A. 2002, A\&A, 390, 1001
\bibitem[Soifer et al.(1989)]{Soifer89} Soifer, B.T., Boehmer, L., Neugebauer, G., \& Sanders, D.B. 1989, AJ, 98, 766
\bibitem[Strickland et al.(2004)]{Strickland04} Strickland, D.K., Heckman, T.M., Colbert, E.J.M., Hoopes, C.G., \& Weaver, K. A. 2004, ApJ, 606, 829
\bibitem[Sutton et al.(1995)]{Sutton95} Sutton, E.C., Peng, R., Danchi, W.C., et al. 1995, ApJS, 97, 455
\bibitem[Takano et al.(2002)]{Takano02} Takano, S., Nakai, N. \& Kawaguchi 2002, PASJ, 54, 195
\bibitem[Viti et al.(2002)]{Viti02} Viti, S., Natarajan, S., \& Williams, D.A. 2002, MNRAS, 336, 797
\bibitem[Wang et al.(2004)]{Wang04} Wang, M., Henkel, C., Chin, Y.-N., et al. 2004, A\&A, 422, 883
\bibitem[Wei\ss\, et al.(2001)]{Weiss01} Wei\ss, A., Neininger, N., Henkel, C., Stutzki, J., \& Klein, U. 2001, ApJ, 554, L143
\bibitem[Weliachew et al.(1984)]{Weli84} Weliachew, L., Fomalont, E.B., \& Greisen, E.W. 1984, A\&A, 137, 335
\end{thebibliography}

\end{document}